\documentstyle%[amssymb,12pt,thmsa,sw20lart]
{elsart}
\def\la{\mathrel{\mathchoice {\vcenter{\offinterlineskip\halign{\hfil
$\displaystyle##$\hfil\cr<\cr\sim\cr}}}
{\vcenter{\offinterlineskip\halign{\hfil$\textstyle##$\hfil\cr
<\cr\sim\cr}}}
{\vcenter{\offinterlineskip\halign{\hfil$\scriptstyle##$\hfil\cr
<\cr\sim\cr}}}
{\vcenter{\offinterlineskip\halign{\hfil$\scriptscriptstyle##$\hfil\cr
<\cr\sim\cr}}}}}
\def\ga{\mathrel{\mathchoice {\vcenter{\offinterlineskip\halign{\hfil
$\displaystyle##$\hfil\cr>\cr\sim\cr}}}
{\vcenter{\offinterlineskip\halign{\hfil$\textstyle##$\hfil\cr
>\cr\sim\cr}}}
{\vcenter{\offinterlineskip\halign{\hfil$\scriptstyle##$\hfil\cr
>\cr\sim\cr}}}
{\vcenter{\offinterlineskip\halign{\hfil$\scriptscriptstyle##$\hfil\cr
>\cr\sim\cr}}}}}
\input{epsf.tex}

\begin{document}

\begin{frontmatter}
\title{The angular power spectra of polarized Galactic synchrotron}
\author [milano,mi2]{M. Tucci\thanksref{rfemail1}},
\author [boh]{E. Carretti\thanksref{rfemail2}},
\author [boh]{S. Cecchini\thanksref{rfemail3}},
\author [unifi]{R. Fabbri\thanksref{rfemail4}},
\author [boh,bo2]{M. Orsini\thanksref{rfemail5}},
\author [ubc]{E. Pierpaoli\thanksref{rfemail6}}
\address [milano]{Dipartimento di Fisica G. Occhialini, Universit\`a
di Milano, Bicocca}
\address [mi2]{I.N.F.N., Sezione di Milano,
Via Celoria 16, I-20133 Milano, Italy}              
\address [boh] {I.Te.S.R.E.-C.N.R., Via P. Gobetti 101,
I-40129 Bologna, Italy}      
\address [unifi]{Dipartimento di Fisica, Universit\`a di Firenze,  
Via S.~Marta 3, I-50139 Firenze, Italy}         
\address [bo2] {Dipartimento di Astronomia, Universit\`a di Bologna,
Via Ranzani 1, I-40127 Bologna, Italy}              
\address [ubc] {Department of Physics and Astronomy, University of
British Columbia, 2219 Main Mall, Vancouver, B.C. V6T 1Z4, Canada}

\thanks[rfemail1]{E-mail: marco@axrialto.uni.mi.astro.it}                                 
\thanks[rfemail2]{E-mail: carretti@tesre.bo.cnr.it}                                   
\thanks[rfemail3]{E-mail: cecchini@bo.infn.it}                               
\thanks[rfemail4]{E-mail: fabbri@dffs.unifi.it;fabbri@fi.infn.it}                                
\thanks[rfemail5]{E-mail: orsini@tesre.bo.cnr.it}                                   
\thanks[rfemail6]{E-mail: elena@astro.ubc.ca}                                                
%\maketitle

\begin{abstract}
We derive the angular power spectra of intensity and polarization of
Galactic synchrotron emission in the range $36\leq \ell \la  10^3$ from
the Parkes survey mapping the southern Galactic plane at 2.4 GHz.
The polarization spectra of both electric and magnetic parity
up to $\ell \simeq 10^3$ are approximated very well by power laws with
slope coefficients $\simeq 1.4$, quite different from the CMB spectra.
We show that no problem should arise from Galactic synchrotron
for measurements of CMB polarization in the cosmological window.
\end{abstract}

\end{frontmatter}

\section{Introduction and main results}

The purpose of this work is to compute the angular power spectra of
synchrotron emission from the Galaxy, and in particular of its polarized
component. Synchrotron radiation is expected to dominate the linearly
polarized component of the sky background in a wide frequency range up to
a few tens of GHz. It is important for the knowledge 
of the Galactic structure, and
also because it will be a major contaminant in forthcoming measurements of
the polarization of the cosmic microwave background (CMB), planned from
ground  (Keating et al., 1998; Sironi et al., 1997; 
Pisano, 1999; Hedman, 1999) 
%\cite{keating,sironi,pisano,hedman} 
and from space (Wright, 1999; Cortiglioni et al., 1999; 
De Zotti et al., 1999).
%\cite{wribenn,cortig,dezber}

Separation of the various contributions to the total background in the
``cosmological window'' (say, $\simeq 50 - 90$ GHz) 
promises to be a delicate job, and its success will
rely on sensible assumptions about the spectral and spatial properties of
Galactic foregrounds (e.g., Tegmark et al. 2000, Prunet et
al. 2000). %\cite{prusebo,teho}  
In this connection we observe that at frequencies of tens of GHz the
dominant contribution to the angular spectrum may be a Galactic foreground
(synchrotron or dust), the integrated background of unresolved extragalactic
sources, the primordial anisotropy of CMB or a non-primordial anisotropy
(for instance arising from the gravitational lens effect) depending on the
range of the spherical-harmonic index $l$. Therefore the cosmological window
should be understood as a bidimensional region of the ($\nu $, $l$) plane,
which is furthermore different for anisotropy and polarization.
%Although Galactic maps at other frequencies can be
%used as templates for estimates on a pixel basis, one can also conveniently
%use harmonic expansions; in fact, 
For this reason angular power spectra, which provide a
standard tool for CMB analysis, are becoming relatively commonplace also for
foregrounds, although in the latter case they provide an incomplete
description because of the strong phase coherence of Galactic emission.
Intensity (or temperature) power spectra of foregrounds 
are reasonably approximated by
power laws, $C_{I\ell }\propto \ell ^{-\alpha _I}$, with slopes steeper than
CMB. For the latter the scale-invariant adiabatic-wave spectrum predicts $%
\alpha _I^{{\rm CMB}}\simeq 2$ on large scales and power excess at $%
10^2\la  \ell \la  10^3$ because of several acoustic peaks. Both the
large-scale behaviour and the first acoustic peak appear to be confirmed by
experiment (Barreiro, 1999, de Bernardis et al. 2000, Hanany et
al. 2000). %\cite{barreiro,deBernardis,Hanany} 
For synchrotron, values $\alpha _I^{{\rm syn}%
}\simeq 2.5\div 3$ have been derived from the 408-MHz maps of Haslam et al. 
(1981) %\cite{haslam} 
by Tegmark \&\ Efstathiou (1996) and Bouchet et al. (1996), %\cite{teg96,bouch96} 
and $\alpha _I^{{\rm syn}}\simeq 3$ from
the 1420-MHz northern sky survey 
(Reich \&\ Reich, 1986) %\cite{reich2} 
by Bouchet \&\ Gispert (1999). %\cite{bouch99} 
However, 
the analysis of Tenerife patch at the same frequencies provides a nearly
scale-invariant spectrum, $\alpha _I^{{\rm syn}}\simeq 2,$ except near the
resolution limit of the maps 
(Lasenby, 1997). %\cite{lasenby} 
For dust emission Gautier et al. (1992) %\cite{gautier} 
gave $\alpha _I^{{\rm dust}}\simeq 3$ down to the IRAS
resolution of a few arc minutes, and a similar result was found for DIRBE 
(Wright, 1998); %\cite{wright}
from the combined DIRBE and IRAS maps Schlegel et al. (1998)
%\cite{schlegel} 
derive $\alpha _I^{{\rm dust}}\simeq 2.5$. The situation is less
clear for free-free emission: While Kogut et al. (1996) %\cite{kog96} 
find $\alpha
_I^{{\rm FF}}\simeq 3$ correlating COBE-DMR with DIRBE, Veeraraghavan \&
Davies (1997) %\cite{vee97} 
derive $\alpha _I^{{\rm FF}}\simeq 2.3$ at 53 GHz from H$%
\alpha $ maps but with a {\it much lower} normalization. The discrepancy
supports the case for dust emission in the cosmological window, the
mechanism of rotational excitation of small grains having been proposed by
Draine \&\ Lazarian (1998). %\cite{draine}

Angular spectra of polarized emission have been measured so far neither for
CMB nor for Galactic foregrounds. Theoretical modeling for CMB gives (for
the electric-parity polarization excited by density waves in the
standard-model) $\alpha _E^{{\rm CMB}}$ rather close to zero, which means
much smaller spatial correlations 
at large angular scales than for temperature fluctuations. A
corresponding excess of small-scale structure (in comparison to temperature
fluctuations) is found by Prunet et al. (1998) %\cite{prus} 
for polarized dust
emission in modeling based on the HI maps of the Leiden-Dwingloo survey
[see also Sethi et al. (1998) %\cite{sethi} 
and Prunet \&\ Lazarian (1999)%\cite{prun}
]:
Their results are $\alpha _E^{{\rm dust}}\simeq 1.3$ and $\alpha _B^{{\rm %
dust}}\simeq 1.4$ for the electric and magnetic parity spectra, respectively.
The angular power spectrum is therefore steeper for
Galactic dust than for CMB.

Now an important question should be posed: can this result be generalized to
other foregrounds? If so, the separation of CMB polarization signals would
be easier. The present work considers the polarization spectra of Galactic
synchrotron, and can be regarded as complementary to that 
of Prunet et al. (1998)
%\cite{prus}\cite{sethi} 
since polarized free-free emission should not be dominant at any
frequency. Our approach is the analysis of low-frequency maps. Unfortunately
no full-sky survey is available for the polarized component of synchrotron,
the largest coverage being provided by Brouw \&\ Spoelstra (1976) %\cite{spoels} 
at
the expense of a quite sparse sampling. Here we make use of the Parkes
survey of the southern Galactic plane (Duncan et al., 1995, 1997)
%\cite{duncan} 
which, although limited
to a relatively small window of $\simeq 0.39$ sr, provides uniformly covered
maps, not suffering from undersampling. Its frequency of 2.4 GHz probably
allows a reasonable extrapolation to the cosmological window, thanks to the
limited effects expected from Faraday 
%rotation:
%Spoelstra (1984) and Wieringa et al. (1993), respectively using data
%on large scales and on small scale structures, found typical values of
%the order of 5--10 $rad$ $m^{-2}$ for the rotation measure (RM) . At
%frequency of 2.4 GHz such values of RM produce  rotation angles of
%approximately a few degrees, too small to produce a considerable
%depolarization. Although the RM is probably higher in particular
%stuctures along the Galactic Plane, however we may consider that
%Faraday rotation does not affect the global distribution of polarized
%intensity. 
rotation. The typical values of 5--10 rad m$^{-2}$ for
the rotation measure (RM) reported by Spoelstra (1984) and
Wieringa et al (1993) allow to estimate a negligible depolarization 
of few percent at 2.4 GHz. 

The main result of this paper is that the synchrotron
polarization spectra 
up to $\ell \simeq 10^3$ are approximated very well by power laws with
slope coefficients $\alpha _E \simeq \alpha _B \simeq 1.4$, a result very
close to that of Prunet et al. (1998). The intensity spectrum
is also found to have a moderate slope, close to that of polarization
spectra, unless we use alternative maps (available at the
Parkes WEB site\footnote{%
http://www.atnf.csiro.au/database/astro\_data/2.4Gh\_Southern}) 
where Galactic-plane sources have been removed; therefore
this result should not be extrapolated to the whole sky.
In the last section of this paper we provide reasons, suggesting that
the slopes of polarization spectra are likely to apply to the whole sky.
We show that if this assumption is correct, synchrotron emission
cannot be a problem for CMB polarization measurements in the
cosmological window. Otherwise,  
a value $\simeq 1.4$ can
be used as a {\it lower limit} to $\alpha _E$ and  $\alpha _B$;
then the slope differences with respect to the CMB polarization spectra 
would still be larger,
and the CMB polarized signal would prevail even more strongly in the
cosmological window.  

\section{Angular spectra in the Parkes survey}

Figure \ref{syncmaps} shows the sky coverage of the Parkes 2.417 GHz survey
and other available surveys as well. The Parkes survey covers a strip, 127$%
^{\circ }$ long and at least 10$^{\circ }$ wide, with a FWHM resolution of $%
10^{\prime }.4$. Maps are available at the Parkes WEB site for the total
intensity $I$ both before and after subtraction of structures with size $%
\leq 2^{\circ }$, and for the Stokes parameters $Q$ and $U$ with no source
subtraction.

{\begin{figure}
\centerline{\epsfxsize=13cm
\epsfbox{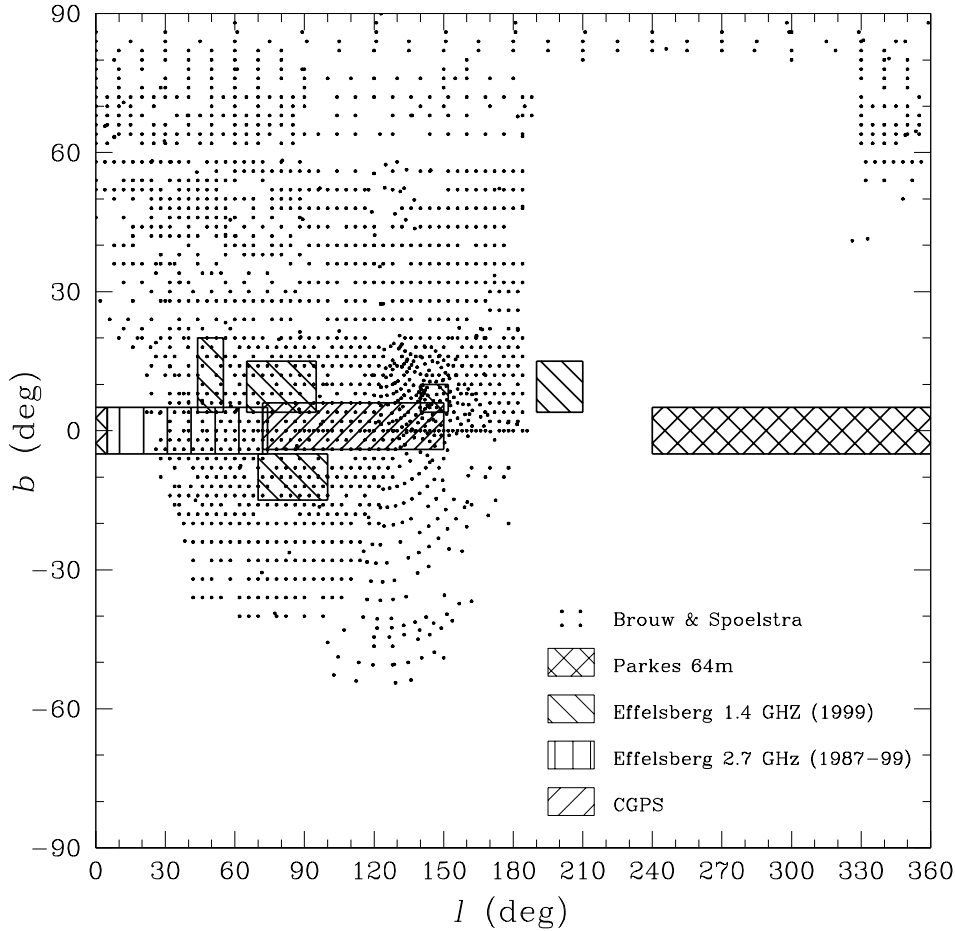}}
\caption{Sky coverages of synchrotron polarization surveys. We
report the regions observed at 0.408-1.411 GHz by
Brouw \&\ Spoelstra (1976); in the 2.417-GHz Parkes survey
(Duncan et al., 1995, 1997); at Effelsberg, 2.695 GHz
(Junkes et al. 1987; Duncan et al., 1999) and 1.4 GHz 
(Uyaniker et al. 1999);
and in the Canadian Galactic Plane survey (CGPS, 
0.408 and 1.420 GHz)  (English et al., 1998). }
\label{syncmaps}
\end{figure}}

{\begin{figure}
\centerline{\epsfxsize=12cm
\epsfbox{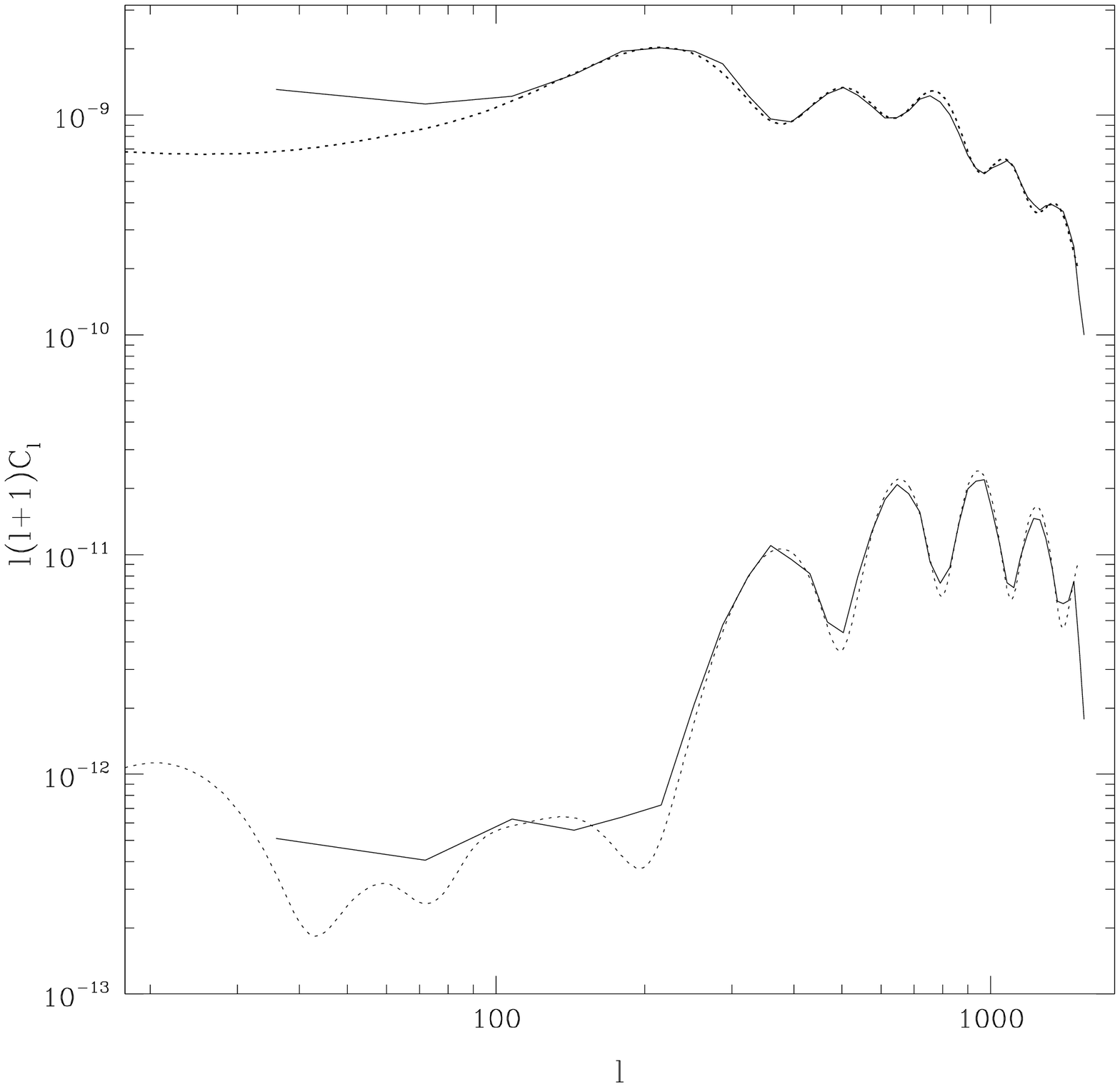}}
\caption{The input and output CMB spectra (dotted and full line, respectively)
for a test of Fourier analysis in the Parkes survey window. The input
spectrum is 
computed for a standard CDM model with a secondary-ionization optical depth
$\tau_{ion}  = 0.18$.}
\label{testcmb}
\end{figure}}

Because of the limited sky coverage, spherical-harmonic spectra are suitably
obtained from a standard Fourier analysis according to the technique
developed by Seljak (1997) %\cite{seljak} 
for the CMB polarization and applied by 
Prunet et al. (1998)
%\cite{prus} 
to the polarized dust foreground. After
the Fourier components of Stokes parameters $I({\bf l})$, $Q({\bf l})$ and $%
U({\bf l})$ are computed on a map covering a solid angle $\Omega $, the
estimators for the power spectra of intensity $C_{I\ell }$ and total
polarization $C_{P\ell }$ can be readily computed by means of the equations 
\begin{eqnarray}
C_{I\ell } &=&\left\{ \frac \Omega {N_\ell }\sum_{{\bf l}}\left[ I({\bf l}%
)I^{*}({\bf l})\right] -w_I^{-1}\right\} b^{-2}(\ell ),  \label{intens} \\
C_{P\ell } &=&\left\{ \frac \Omega {N_\ell }\sum_{{\bf l}}\left[ Q({\bf l}%
)Q^{*}({\bf l})+U({\bf l})U^{*}({\bf l})\right] -2w_P^{-1}\right\}
b^{-2}(\ell ),  \label{polar}
\end{eqnarray}
where the sums are performed over the $N_\ell $ modes with wavevector
magnitude around $\ell $, $b(\ell )$ is the window function and $%
w_{I,P}^{-1}=\Omega \sigma _{I,P}^2/N_{{\rm pixel}}$ is the
pixel-independent measure of noise with $\sigma _{I,P}$ the pixel 
noise. The electric and magnetic parity of
polarization can also be separately computed by means of 
\begin{eqnarray}
C_{E\ell } &=&\left\{ \frac \Omega {N_\ell }\sum_{{\bf l}}\left| Q({\bf l}%
)\cos (2\phi _{{\bf l}})+U({\bf l})\sin (2\phi _{{\bf l}})\right|
^2-w_P^{-1}\right\} b^{-2}(\ell ),  \label{electr} \\
C_{B\ell } &=&\left\{ \frac \Omega {N_\ell }\sum_{{\bf l}}\left| -Q({\bf l}%
)\sin (2\phi _{{\bf l}})+U({\bf l})\cos (2\phi _{{\bf l}})\right|
^2-w_P^{-1}\right\} b^{-2}(\ell ),  \label{magnet}
\end{eqnarray}
with $\phi _{{\bf l}}$ the direction angle of ${\bf l}.$

For the Parkes survey the window function can be approximated by a Gaussian, 
$b(\ell )=\exp \left[ -\ell (\ell +1)\sigma _{{\rm b}}^2/2\right] $ with $%
\sigma _{{\rm b}}=4^{\prime }.4$ . The polarization noise is not constant,
being 11 mJy beam area$^{-1}$ for most of the sky coverage but as low as 6
mJy beam area$^{-1}$ in some regions. We chose to extract from the survey
constant-noise square patches of $10^{\circ } \times 10^{\circ }$ (i.e.,
$150 \times 150$ pixels). We thereby obtained
12 independent submaps (the first one being centered at
$l = 360^{\circ}$ and moving toward decreasing Galactic longitudes),
and on each of them we performed the Fourier
analysis and derived a set of four spectra $C_{X\ell }$ (with $X=I$, $P$, $E$
and $B$) by means of Eqs. (\ref{intens})-(\ref{magnet}) (the $P$ spectrum
was in fact redundant, but we computed it for all of the submaps for checks
of consistency). 
A potential problem with the Fourier analysis, when it is applied to small
maps, arises from border effects, which may give spurious contributions
to the spectra. A solution to this problem is considered by 
Hobson \&\ Magueijo (1996),
who suggest to modulate the signal by some function vanishing
at the borders (apodization). 
We carried out our analysis both with appropriate cosine functions
and with no modulation. The differences turned out to be small in the
range $\ell \la 1000$, except for the intensity maps
with source subtraction. In Figures \ref{twelvepatch}-\ref{meanspec1}
we explicitly report the results
obtained with the cosine modulation up to $\ell = 1100$, which is
close to the resolution limit of the survey.
In order to test the reliability of the computed spectra,
we also built up simulated maps for CMB, using angular spectra
computed by means 
of CMBFAST (Seljak \&\ Zaldarriaga, 1996) %\cite{seljak2}) 
and HEALPix package\footnote{http://www.tac.dk/~healpix},
and performed the Fourier analysis on them
within the Parkes survey window. The results shown in Figure \ref{testcmb}
prove that the method is quite reliable: The output spectrum differs from
the input one only for some smoothing of the sharpest features,
which is particularly evident for low values of $\ell $. 
Smoother spectra -- such as those expected for synchrotron -- are less
affected by the procedure. The results
for all of the 12 submaps also are mutually consistent in the CMB case.

{\begin{figure}
\centerline{\epsfxsize=12cm
\epsfbox{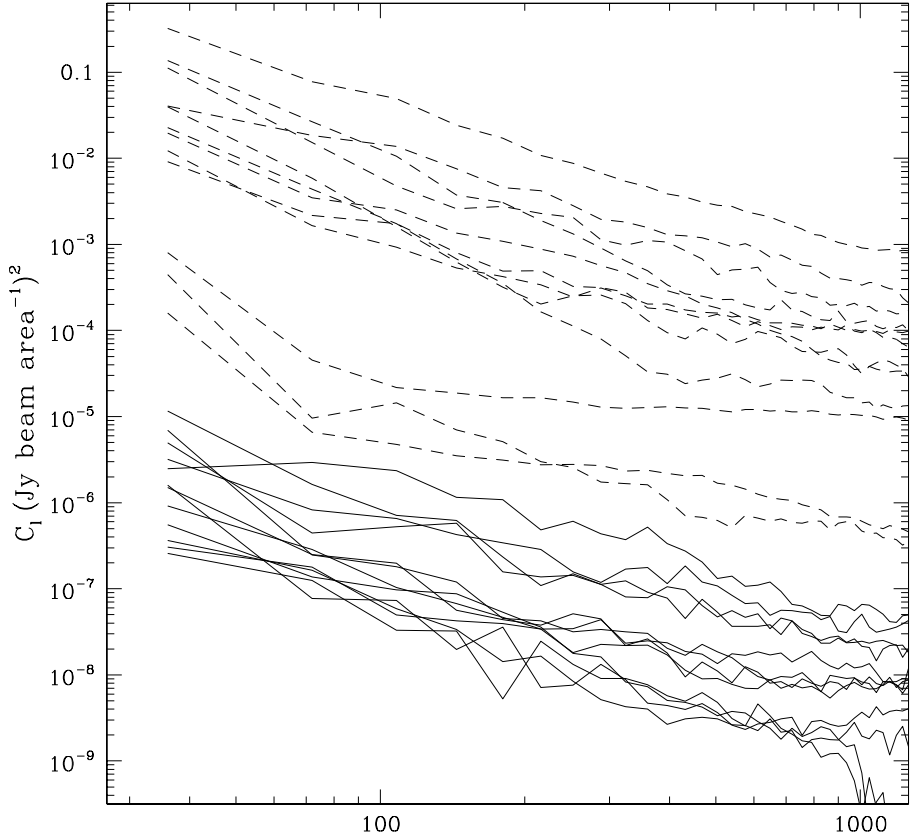}}
\caption{The angular power spectra $C_{I \ell } $  and
$C_{E \ell }$ (dashed and full lines, respectively) for the 12 submaps of the
Parkes survey.}
\label{twelvepatch}
\end{figure}}

{\begin{figure}
\centerline{\epsfxsize=12cm
\epsfbox{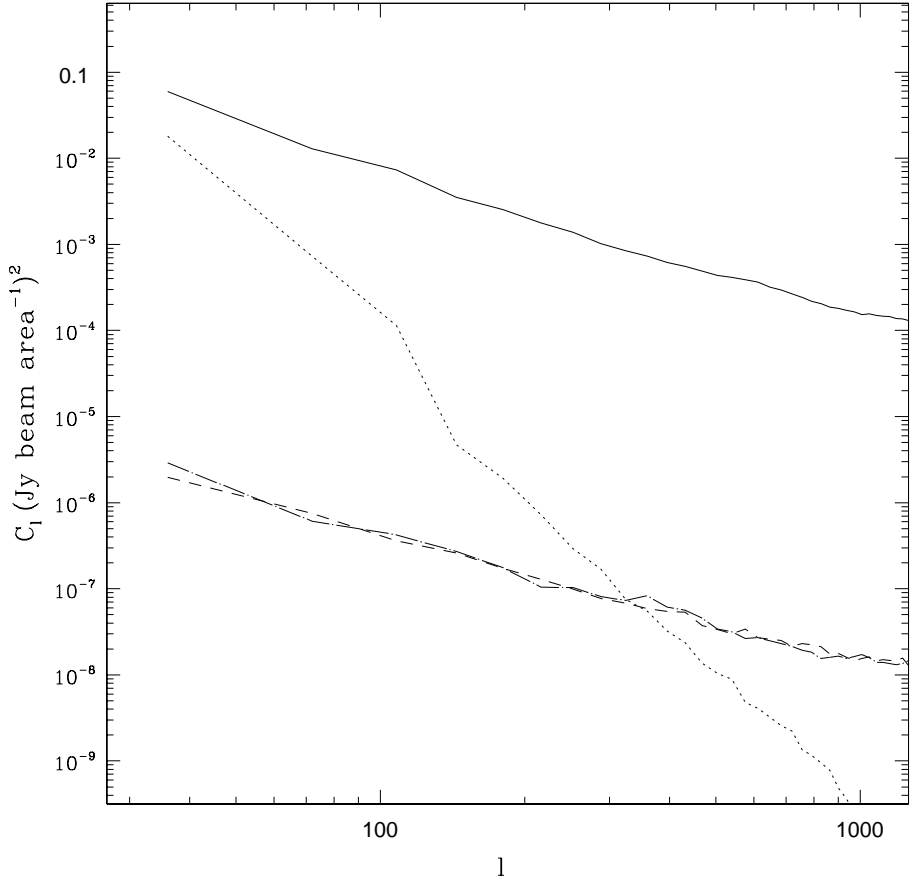}}
\caption{The mean
power spectra in the Parkes survey. The curves give the intensity 
$C_{I \ell } $ with
and without source subtraction (full and dotted lines, respectively)
and the polarization $C_{E \ell }$ and  $C_{B \ell }$  
(dash-dotted and dashed).}
\label{meanspec0}
\end{figure}}

{\begin{figure}
\centerline{\epsfxsize=12cm
\epsfbox{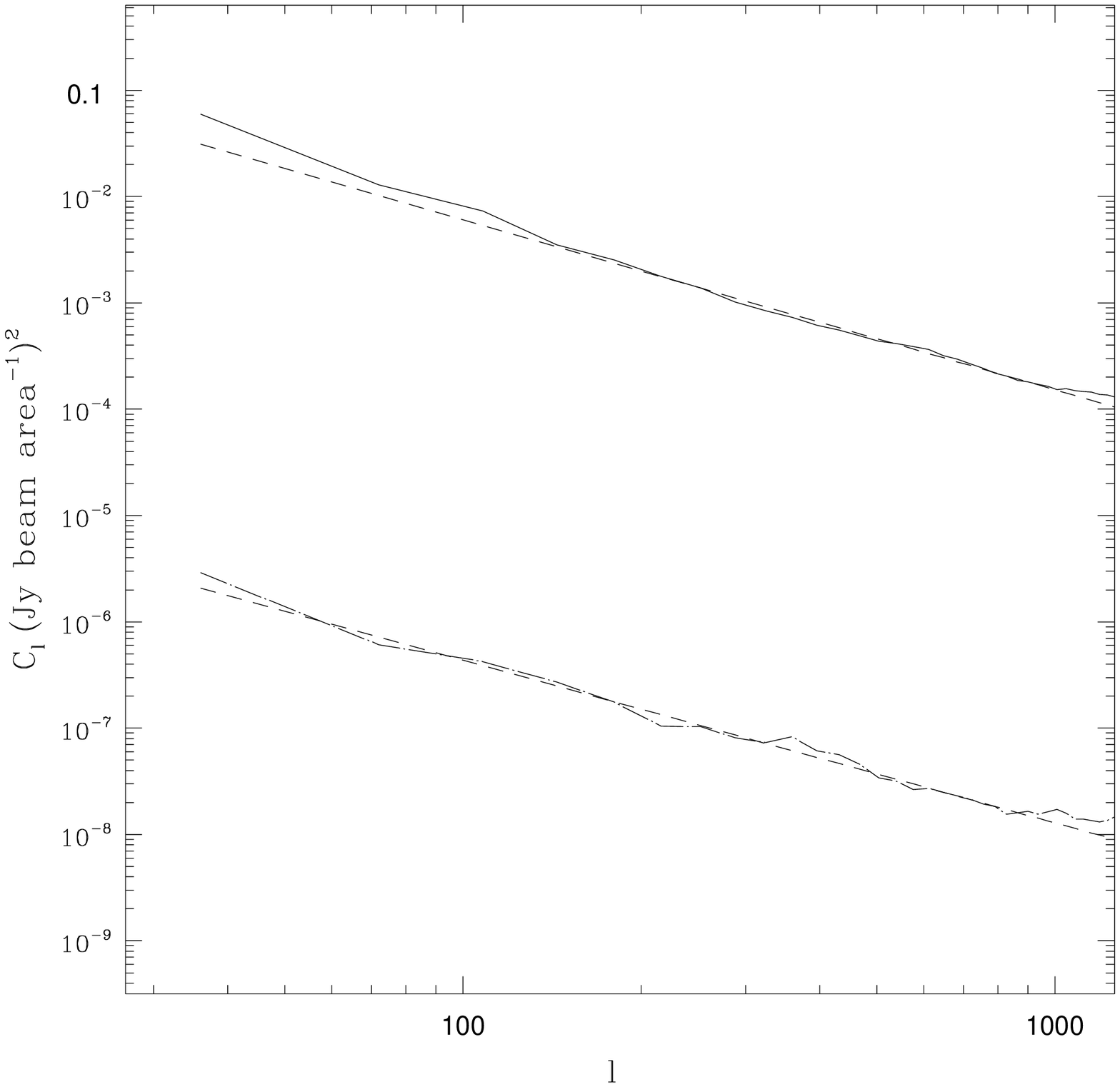}}
\caption{The intensity and electric parity spectra compared to the 
fitting functions described in the text.
The best-fit parameters refer to the range $36 \le \ell \le 800$
and to averaging ``method 2''.}
\label{meanspec1}
\end{figure}}

The spectra obtained from synchrotron submaps, however, 
show large variations in the normalization; this fact is
certainly not unexpected, due to the phase coherence of Galactic structure.
Figure \ref{twelvepatch} shows the results for $C_{I\ell }$ (with no source
subtraction) and $C_{E\ell }$ from the individual submaps; the results for
magnetic parity polarization are similar to those for electric parity
(the intensity spectra after source removal, however, are much steeper than those
reported in the Figure, as we will discuss below).

The normalization variations for intensity and polarization cover different
ranges (spanning more than 3 and about 2 orders of magnitude, respectively), and
they are not strictly correlated since the polarization degree is far from
being constant. The slopes of the curves up to $\ell \approx 10^3,$ however,
are much more mutually consistent than normalization. Visual inspection of
Fig. \ref{twelvepatch} shows a couple of low-emission regions
(submaps 9 and 12, placed at $ l < 280^{\circ }$) with
particularly flat intensity curves
at $\ell \ga 10^2$, but this feature does not appear in
polarized emission. Clearly most of the curves 
are reasonably approximated by power
laws, but when we fit the data with functions $C_{X\ell }=A_X\ell ^{-\alpha _X}$, 
the logarithmic slopes vary with both the chosen submap
and the range of $\ell $.
In Table \ref{slopes1} we report angular-spectrum
best-fit slopes for all of the submaps with no source subtraction. 
Columns 2 to 4 (primed quantities) refer to the range 
$\ell \leq 500$, and the remaining (unprimed quantities) to 
$\ell \leq 800$. Only for two submaps and for
$\ell \leq 500$ the intensity spectra are steeper than
$\alpha _I = 2$; more often they are only moderately steeper
than polarization spectra, or even hardly
distinguishable from them (as far as slopes are concerned).

{\begin{table}
\caption[Best-fit slopes of angular power spectra in the Parkes survey submaps 
with no source subtraction]%
{Best-fit slopes of angular power spectra in the Parkes
survey submaps 
with no source subtraction 
\label{slopes1}}
\begin{tabular}{ccccccc}
\hline
Submap
& $\alpha_I^{\prime}$ &  $\alpha_E^{\prime}$ &  $\alpha_B^{\prime}$    
& $\alpha_I$ &  $\alpha_E$ & $\alpha_B$   \\
\hline
 1 & 1.79 & 1.05 & 1.24 & 1.65 & 1.28 & 1.33 \\
 2 & 1.50 & 1.28 & 1.25 & 1.64 & 0.85 & 1.12 \\
 3 & 2.54 & 1.29 & 1.45 & 2.00 & 1.31 & 1.52 \\
 4 & 1.98 & 1.74 & 2.06 & 1.46 & 1.57 & 1.74 \\
 5 & 2.87 & 1.56 & 1.47 & 1.76 & 1.49 & 1.40 \\
 6 & 1.65 & 1.16 & 1.18 & 1.89 & 1.11 & 0.96 \\
 7 & 1.56 & 1.39 & 1.78 & 1.22 & 1.56 & 1.88 \\
 8 & 1.56 & 2.26 & 1.98 & 1.43 & 2.04 & 1.90 \\
 9 & 0.74 & 1.78 & 2.02 & 0.44 & 1.56 & 1.78 \\
10 & 1.23 & 1.28 & 1.23 & 1.02 & 1.79 & 1.25 \\
11 & 1.97 & 1.83 & 1.61 & 1.48 & 1.48 & 1.32 \\
12 & 0.76 & 2.09 & 1.68 & 0.96 & 1.58 & 1.45 \\ \hline
\end{tabular}
\end{table}}

\begin{table}
\caption[Best parameters for angular power spectra]%
{Best parameters for angular power spectra
\label{slopes2}}
\begin{tabular}{lllllll}
\hline
$X$ & Averaging  & $A_X^{\prime }$ & $\alpha _X^{\prime}$ & $A_X$ & $\alpha _X$ \\
& method & Jy (beam area)$^{-1}$ &  & Jy (beam area)$^{-1}$ &  \\ \hline
$I$ & 1 & $241\pm 109$ & $1.71\pm 0.18$ & $17\pm 9$ & $1.37\pm 0.13$ \\ 
$I$ & 2 & $28^{+27}_{-13}$ & $1.79\pm 0.13$ & $9.6^{+11}_{-5}$ & 
$1.6\pm 0.13$ \\
$I$ & $2^{*}$ & $(7.5_{-5.3}^{+25})\cdot 10^{+5}$ & $5.2\pm 0.3$ &
$(2.5_{-1.6}^{+2.3})\cdot 10^{+5}$ & $5.0\pm 0.2$ \\
$I$ & 3 & $12^{+22}_{-9}$ & $1.6\pm 0.2$ & $3.2^{+3.5}_{-1.8}$ &
$1.36\pm 0.13$ \\
$E$ & 1 & $(8.5\pm 3.8)\cdot 10^{-4}$ & $1.57\pm 0.11$ & $(11.5\pm
8.7)\cdot 10^{-4}$ & $1.44\pm 0.09$  \\ 
$E$ & 2 & $(2.1^{+4.7}_{-1.6})\cdot 10^{-4}$ & $1.37\pm 0.22$ & 
$(5.0^{+4.4}_{-2.5})\cdot 10^{-4}$ & $1.53\pm 0.11$ \\
$E$ & 3 & $(1.3^{+2.2}_{-0.9})\cdot 10^{-4}$ & $1.31\pm 0.18$ & $(2.7\pm
1.2)\cdot 10^{-4}$ & $1.45\pm 0.1$ \\
$B$ & 1 & $(16\pm 12)\cdot 10^{-4}$  & $1.58\pm 0.10$ & $(4.4\pm
1.8)\cdot 10^{-4}$ & $1.46\pm 0.09$  \\ 
$B$ & 2 & $(4.5^{+8}_{-3})\cdot 10^{-4}$ & $1.52^{+0.16}_{-0.20}$ & 
$(2.8^{+3.1}_{-0.8})\cdot 10^{-4}$ & $1.43^{+0.13}_{-0.06}$  \\
$B$ & 3 & $(8.2^{+14}_{5.6})\cdot 10^{-4}$ & $1.35\pm 0.20$ &
$(6.8^{+6.7}_{-3.9})\cdot 10^{-4}$ & $1.32\pm 0.13$  \\ \hline
\end{tabular}
\begin{minipage}[t]{4.in}
$^{*}$After source subtraction
\end{minipage}
\end{table}

We then have to combine the results from the 12 submaps in some way.
A simple method is to compute weighted averages of the best-fit
parameters; this implies weighting with statistical
errors, with no regard to the strength of the emission in the submaps.
Table \ref{slopes2} gives the mean parameters (including the normalization
factors, denoted by $A_X$ and $A_X^{\prime }$) obtained in this way, 
as well as the statistical errors; they are collected in the Table rows
referring to averaging ``method 1''. According to such results, 
the slope of the intensity spectrum slightly decreases for increasing
$\ell $ and can be hardly distinguished from that of polarization spectra 
at large $\ell $: 
\begin{equation}
\alpha _I \simeq \alpha _E \simeq  \alpha _B \simeq 1.4 \div 1.5,
\; \; (\ell = 36 \div 800,\; \mathrm{method}\; 1).
\label{method1}
\end{equation} 
Similar results are found with the simpler treatment with no signal
cosine modulation; the differences with respect to the results in the
Table are $\Delta \alpha _X \simeq \pm 0.1$.

It can be argued that because of the averaging method, the slopes given by
Eq. (\ref{method1}) may be
not representative of the full Parkes
survey, since low-emission and high-emission subwindows are considered on
equal footing. An alternative approach is
averaging the spectra of the 12 subwindows:
Figure \ref{meanspec0} reports the mean spectra, 
and includes magnetic parity and the
intensity spectrum after source subtraction. 
We thereby performed best fits as above also on such mean
spectra. This procedure, referred to as averaging
``method 2'' in Table \ref{slopes2}, clearly gives stronger 
weights to high-emission
submaps.
The results are quite different for the normalization factors 
(and especially for intensity), but no large variation is found for the
slope factors. The slope of the intensity spectrum however
is slightly increased now, 
due to the scarce weight given to the nearly flat spectra of a couple
of low-intensity submaps, the larger value being found
in the smaller range $\ell = 36 \div 500$; no clear change 
however is found for
polarization:

\begin{equation}
\alpha ^{\prime} _I \simeq 1.8, \; \; \alpha ^{\prime}_E \simeq  
\alpha ^{\prime}_B \simeq 1.4 \div 1.5,
\; \; (\ell = 36 \div 500,\; \mathrm{method}\; 2).
\label{method2}
\end{equation}  

As shown by Fig. \ref{meanspec1}, the fitting functions with the
parameters found  in the larger range ($\ell \le 800$) describe 
very well the polarization spectra, whereas a modest discrepancy
appears for the intensity spectrum at $\ell \le 100$.

We also tried with a further procedure, 
renormalizing the spectra of the 12 submaps to a same mean intensity value
and then constructing mean curves. The parameters of such mean curves are
referred to as averaging ``method 3'' in the Table, and show slightly
smaller slopes. For
the range $\ell = 36 \div 800$ 
\begin{equation}
\alpha _I \simeq  \alpha _E \simeq  \alpha _B \simeq 1.4,  \; \;
(\ell = 36 \div 800,\; \mathrm{method}\; 3).
\label{method3}
\end{equation}  

A quite different result is found for the intensity maps with source
subtraction. The slope is now much higher, $\alpha _I \simeq  5$. At
large $\ell$ the signal is very small, and it would be overcome by border
effects in the absence of apodization even for $\ell < 10^3$.

\section{Discussion}

The most attractive conclusion we can draw from the present work regards the
angular spectrum slope of the polarized synchrotron emission up to $\ell
\approx 10^3$. Since this slope substantially differs from that of polarized
CMB, provided it can be extrapolated to the cosmological window the
separation of cosmological and Galactic signals 
becomes easier. In
particular, at a given frequency CMB polarization is expected to prevail
better at smaller scales; this result is derived from Galactic-plane maps
with no source subtraction. It is interesting to observe that essentially
the same slope, $\alpha _E\simeq \alpha _B\simeq 1.4,$ was found by Prunet et
al. (1998) %\cite{prus} 
for polarized dust emission. We may speculate that this coincidence
may be due to the similarity of polarizing/depolarizing 
mechanisms inscribed in the
Galactic structure.

{\begin{figure}
\centerline{\epsfxsize=12cm
\epsfbox{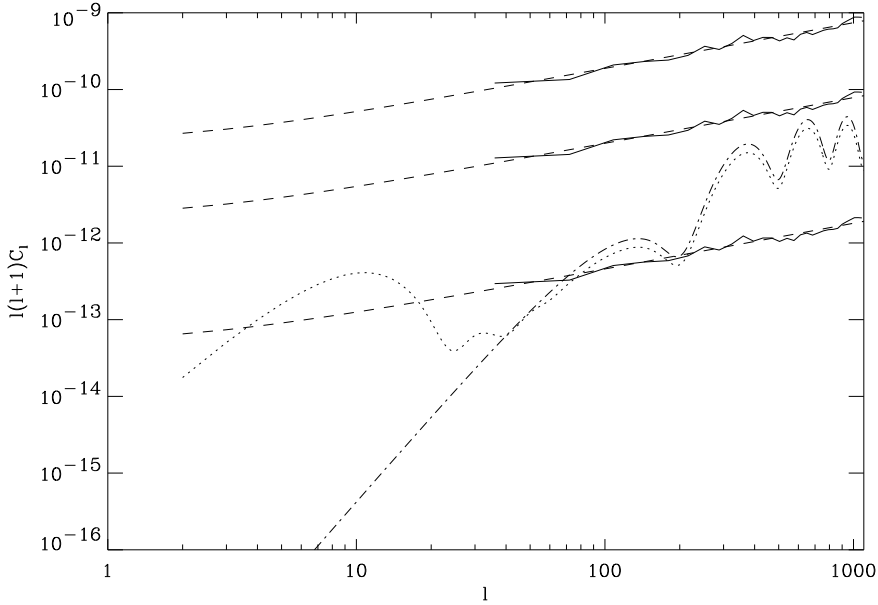}}
\caption{The angular spectra of polarized synchrotron extrapolated
to 22, 32, and 60 GHz (full line, from top to down) with the linear
fits (dashed line), compared to the CMB E-spectrum computed for
a standard CDM model (dot-dashed line) and a CDM with a secondary 
ionization optical depth
$\tau _{ion} = 0.2$ (dotted line).}
\label{synchcmb}
\end{figure}}

On the other hand, our results on the total intensity spectrum might
look somewhat intriguing at first
sight. The original Parkes maps with no source subtraction support only
a moderate slope, close to that of polarization spectra. 
This contrasts sharply with the
results of  Tegmark \&\ Efstathiou (1996), Bouchet et al. (1996) and Bouchet
\&\ Gispert (1999);
%\cite{teg96,bouch96,bouch99} 
the discrepancy is much smaller with the analysis of 
Lasenby (1997) %\cite{lasenby} 
on the Tenerife patch. Removal of sources up to $2^{\circ }$ wide,
however, gives a much larger slope, beyond the more 
familiar result $\alpha _I\approx 3$. This fact can be readily interpreted,
observing that in the Parkes survey
the intensity angular spectrum at large $\ell $ is dominated by Galactic-plane
sources. An obvious consequence is that no real contradiction exists with 
previous work, since such results as $\alpha _I \simeq 1.4 \div 1.8$
cannot be extrapolated outside the
Galactic plane.

We then should ask, which values of $ \alpha _X $ 
would be appropriate for the whole sky. 
Clearly the subtraction of $2^{\circ }$ sources as described
at the Parkes WEB site is a quite radical procedure, cutting out 
high-order harmonics very efficiently. It
is then easy to infer that the full-sky $ \alpha _I $ 
must lie somewhere
between 1.4 and 5; however the existing support for the value
$\alpha _I\approx 3$ is not very strong, in view of
the results of  Lasenby (1997), supporting $\alpha _I\approx 2.$
On the other hand, we believe that our results on the polarization slope 
can be extrapolated out of the Galactic plane
with some confidence. The inspection of Fig.
\ref{twelvepatch} shows that for polarization 
no systematic differences of slope exist
between high and low emission submaps. As already remarked,
the overall normalization (rms signal)
has much weaker variations than for intensity spectra. 
Thus the role of Galactic-plane 
sources
is less important. Also, the agreement with Prunet et al. (1998) is
impressive.

Assuming that the polarization angular spectra are correct for the full sky, 
we can try to extrapolate them to frequencies higher that 2.4 GHz and compare
them to CMB spectra. Figure \ref{synchcmb} shows such extrapolations to
22, 32 and 60 GHz, where the average
Parkes survey normalization and a synchrotron spectral index of 3 are 
assumed. Obviously this procedure is expected to overrate the synchrotron
full-sky spectra. The comparison of synchrotron and CMB angular
spectra shows that for $\ell > 10^2$ the latter prevail
at 60 GHz in spite of the above Galactic-plane normalization.

Finally, we may ask how this conclusion should be changed if the
polarization slopes of Table \ref{slopes2} cannot apply to a
full-sky survey. According to the reasoning expounded for intensity spectra,
we can state that values $\simeq 1.4$ can
{\it at least} be used as lower limits to $\alpha _E$ and  $\alpha _B$.
Then the slope differences with respect to the 
CMB polarization spectra would still
increase in comparison with the scenario described by Fig. \ref{synchcmb},
and the CMB signal would prevail even more strongly in the
cosmological window.  

\section*{Acknowledgments}

We thank S. Bonometto and all the people of the SPOrt collaboration
team for help and
encouragement. M. Tucci wishes to thank D. Scott for stimulating discussions.
We acknowledge use of the CMBFAST code and the HEALPix package.
This work is supported by the Italian Space Agency (ASI).

\end{document}